\documentclass[referee]{raa}            % referee version: for submission
\usepackage{graphicx,times}             %for PS/EPS graphics inclusion, new
\usepackage{threeparttable}
\usepackage{subfigure}
\usepackage{float}
\usepackage{epstopdf}

\begin{document}

   \title{Detection of Large Color Variation of Potentially Hazardous Asteroid (297274) 1996 SK
}

   \volnopage{Vol.0 (200x) No.0, 000--000}      %%preserved for Editor. DOn't remove!
   \setcounter{page}{1}          %%starting page, preserved for Editor. DOn't remove!

   \author{Chien-Hsien Lin
      \inst{1}
   \and Wing-Huen Ip
      \inst{1,2}
   \and Zhong-Yi Lin
      \inst{2}
   \and Fumi Yoshida
      \inst{3}
   \and Yu-Chi Cheng
      \inst{2}
   }

   \institute{Space Science Institute, Macau University of Science and Technology, Avenida Wai Long, Taipa, Macau; {\it chlin@must.edu.mo}\\
       \and 
             Institute of Astronomy, National Central University, No. 300, Jhongda Rd, Jhongli City, Taoyuan County, Taiwan\\
       \and
             National Astronomical Observatory, Osawa, Mitaka, Tokyo 181-8588, Japan\\
   }
   
\vs \no
   {\small Received 2013 April 22; accepted 2013 September 5}

\abstract{Low-inclination Near-Earth Asteroid (297274) 1996 SK, which is also classified as a Potentially Hazardous Asteroid, has a highly eccentric orbit. It was studied by multi-wavelength photometry within the framework of an NEA color survey at the Lulin Observatory. We report here the finding of large color variation across the surface of (297274) 1996 SK within one asteroidal rotation period of $4.656\pm0.122$ hours and classify it as a S-type asteroid according to its average colors of $B-V=0.767\pm0.033$, $V-R=0.482\pm0.021$, $V-I=0.801\pm0.025$ and the corresponding relative reflectance spectrum. It might be indicative of differential space weathering effect or compositional inhomogeneity of the surface materials. 
}

   \authorrunning{C.-H. Lin et al. }            %author_head in even pages
   \titlerunning{Detection of Large Color Variation of PHA (297274) 1996 SK}  % title_head in odd pages
   \maketitle

\section{Introduction}           %% first-level sections will be auto-capitalized
\label{sect:intro}

The main asteroid belt located between the orbits of Mars and Jupiter is composed of a population of small bodies of primitive composition. The largest member, (1) Ceres, with a diameter of 914 km, will be visited by the DAWN spacecraft in 2015. Following (1) Ceres, (2) Pallas (544 km), (4) Vesta (525 km) and (10) Hygiea (431 km) are the most massive asteroids, which might be classified as dwarf planets. Smaller objects down to the size of km and sub-km range are mostly ejectas from impact cratering and/or catastrophic fragments via collisional process (Bottke et al.~\cite{Bot02, Bot05}). Yoshida et al.~(\cite{Yos04}) discussed in detail the collisional evolution of asteroid families using the young Karin family as an example. They pointed out that photometric measurements of the asteroid family members could provide important clues to the corresponding orbital evolution, internal composition and surface effects because of space weathering process (Clark et al.~\cite{Cla02}; Sasaki et al.~\cite{Sas01}). 

Because of the long-term gravitational perturbations of Jupiter and Saturn, some of the collisional fragments could be injected into orbits intercepting the orbits of the terrestrial planets, which potentially cause surface impact events. These scattered stray bodies are further classified as the Amor asteroids if their perihelion distances (q) are between 1.3 AU and 1.017 AU, the Apollo asteroids if their semi-major axis $a> 1.0$ AU and $q<1.017$ AU, and Aten asteroids if $a<1.0$ AU and the aphelion $Q>0.983$ AU. As shown by Bottke et al.~\cite{Bot02}, the majority of these terrestrial planets-crossing asteroids is from the inner asteroid belt even though some of them could be originated from the middle or outer asteroid belt or of cometary origin. 

Among the Near-Earth asteroids (NEAs), which are the general term for the Apollo and Aten asteroids, a number of them have non-zero probability of hitting the Earth in future. For example, it has been estimated that the total number of a subgroup of NEAs called Potentially Hazardous Asteroids or PHAs with $D>100$ m is approximately 4700$\pm$1450 (Mainzer et al.~\cite{Mai12}). A close monitoring and in-depth investigations of the basic physical properties of the PHAs like sizes, shapes and compositions are therefore important. In addition, the PHAs could also represent very valuable natural resources for space exploration and utilization because of their relatively easy access. With these key issues in mind, we have initiated a cooperative project at the Space Science Institute, Macau University of Science and Technology together with the Astronomy Institute, National Central University, to produce a photometric survey of the taxonomical types of NEAs in low inclination orbits. In this work we report the results of an interesting object (297274) 1996 SK, an Apollo asteroid and PHA, based on the observations on May 22 and 23, 2012, at Lulin Observatory, Taiwan. The observations are described in Section 2. The results of the data analysis are given in Section 3. In Section 4, a summary and discussion on the implications of the physical properties of the color variation will be given.

\section{Observations}
\label{sect:Obs}

In our first set of observational targets, the selection criteria are (1) their lack of prior measurements of the lightcurves and surface color, and (2) the suitability of their optical brightness for time-series photometry. Asteroid (297274) 1996 SK of absolute magnitude $H_v = 16.866$, with a semi-major axis $a = 2.434$ AU, eccentricity $e = 0.794$ and inclination $i = 1.962^\circ$ was close to opposition and satisfied these condition in May, 2012. With its perihelion distance $q = 0.5$ AU and low inclination, (297274) 1996 SK is classified as a PHA. It was observed on May 22 and 23, 2012 by multi-filter photometry using the LOT, one-meter telescope at Lulin Observatory. The CCD imaging camera is the PI-1300B, which has 1340 x 1300 pixels with effective pixel scale of $0.516''$. 

The observational log is given in Table~\ref{tab1}. The filters used are broad-band Bessel $BVRI$, which have centered wavelengths of 442, 540, 647, and 786 nm, respectively. The R-band exposure time is 60 seconds per frame and the measurement sequence was made of continuous 20 frames for each run. In total, seven runs were made. However, due to the unstable weather on May 23, much fewer data were acquired. Three sets of B, V and I filter measurements were made in the first half night of May 22, and another set was made in the next night. The Landolt standard star fields used for color calibration were SA107 on May 22 and SA109 on May 23 (Landolt~\cite{Lan92}). The calibrated absolute magnitudes and colors of each star are listed in Table~\ref{tab2}. The photometric accuracy is 0.044 on average. All targets were observed with airmass of ＜ 2 during the nights. 

     \begin{table}
     \begin{center}
       \begin{threeparttable}
         \caption{Observation Log of (297274) 1996 SK}\label{tab1}
         %\begin{minipage}{0.10\linewidth}% Added
         \begin{tabular}{cccccccc} 
           \hline 
           Instrument&	Filter&	Exposure& Date&	$r^*$& $\Delta^*$& $\Phi^*(^\circ)$& Airmass\\ 
           \hline 
&  &~60s/frame&~May 22&~1.454&~0.443&~4.218&~1.28-1.97\\[-1ex]
\raisebox{1.5ex}{LOT} & \raisebox{1.5ex}{$B,V,R,I$} &~60s/frame&~May 23&~1.467&~0.456&~4.987&~1.28-1.53\\
           \hline  
         \end{tabular}
         %\end{minipage}% Added
         %\hfill% Added
         %\begin{minipage}{0.30\linewidth}% Added
         \begin{tablenotes}
        \item *: The quantity on 16:00 UT of each date;
        \item r: Heliocentric distance (A.U.);
        \item $\Delta$: Geocentric distance (A.U.);
        \item $\Phi$: The phase angle of Sun-target-observer.
         \end{tablenotes}
         %\end{minipage}% Added
       \end{threeparttable}
       \end{center}
     \end{table}

\begin{table}
\begin{center}
\begin{threeparttable}
\caption[]{Mean Calibrated Absolute V-band Magnitudes and Colors of Landolt Standard Stars Observed on May 22 and 23\label{tab2}}
 \begin{tabular}{ccccccccc}
  \hline\noalign{\smallskip}
Star & $V^\alpha$ &$V^\beta$ &$(B-V)^\alpha$ &$(B-V)^\beta$ &$(V-R)^\alpha$ &$(V-R)^\beta$ &$(V-I)^\alpha$ &$(V-I)^\beta$\\
  \hline\noalign{\smallskip}
107 459	&12.284	&12.252	&0.900	&0.915	&0.525	&0.370	&1.045	&0.940\\
107 457	&14.910	&14.887	&0.792	&0.830	&0.494	&0.507	&0.964	&0.971\\
107 456	&12.919	&12.875	&0.921	&0.918	&0.537	&0.549	&1.015	&1.035\\
107 592	&11.847	&11.895	&1.318	&1.204	&0.709	&0.389	&1.357	&1.050\\
107 599	&14.675	&14.671	&0.698	&0.727	&0.433	&0.463	&0.869	&0.898\\
107 600	&14.884	&14.863	&0.503	&0.540	&0.339	&0.358	&0.700	&0.715\\
107 601	&14.646	&14.632	&1.412	&1.441	&0.923	&0.949	&1.761	&1.787\\
107 602	&12.116	&12.116	&0.991	&0.934	&0.545	&0.367	&1.074	&0.962\\
  \hline\noalign{\smallskip}
109 949	&12.828	&12.829	&0.806	&0.805	&0.500	&0.503	&1.020	&1.024\\
109 954	&12.436	&12.435	&1.296	&1.305	&0.764	&0.756	&1.496	&1.491\\
109 956	&14.639	&14.644	&1.283	&1.269	&0.779	&0.788	&1.525	&1.533\\
  \noalign{\smallskip}\hline
\end{tabular}

%% place \tablecomments and \tablerefs below \end{center| and \end{center}:
%% you may leave the table-width parameter to editors or set to your actual size
   \begin{tablenotes}
    \item $\alpha$: Magnitudes and color indices from Landolt ~\cite{Lan92};
    \item $\beta$: Mean values measured from this study.
   \end{tablenotes}
   \end{threeparttable}
   \end{center}
\end{table}

The standard data processing was performed by using IRAF program (Image Reduction and Analysis Facility, supplied by National Optical Astronomy Observatories) with $ccdproc$ package for image reduction, $apphot$ for photometry, and $photcal$ for standard stars flux calibrations.

   \begin{figure}[b]
   \centering
   \subfigure{\label{fig1a}
   \includegraphics[width=0.63\columnwidth]{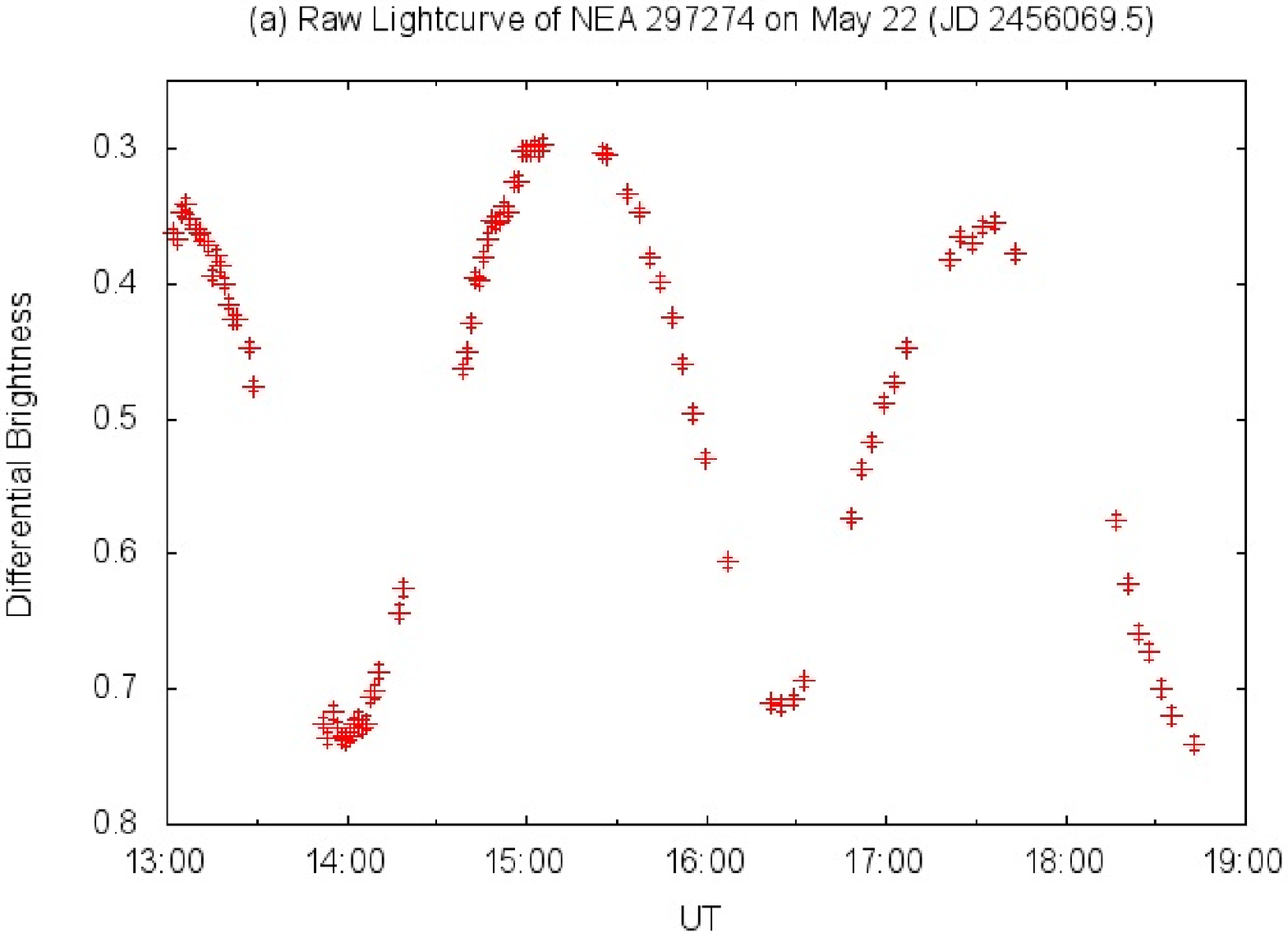}
   }
    \subfigure{\label{fig1b}
   \includegraphics[width=0.33\columnwidth]{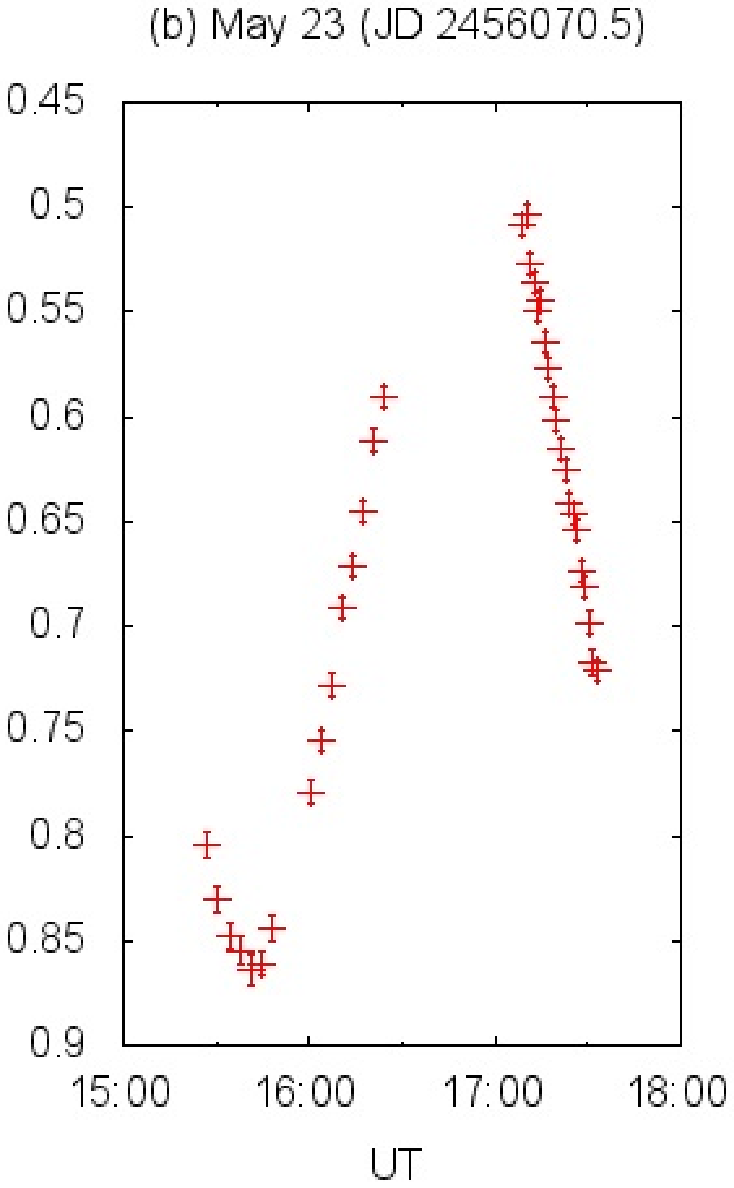}
   }
   
   \caption{The diagrams are respectively the raw lightcurves of (297274) 1996 SK on (a) May 22 across the UT and (b) May 23.}
   \label{Fig1}
   \end{figure}

\section{Results}
\label{sect:results}

Figure~\ref{Fig1} shows the raw lightcurves of (297274) 1996 SK observed on May 22 and 23. Differential photometry was applied while the reference stars without time variability were chosen with R-band magnitude brighter than $17.0$ in the USNO-A2.0 catalog.

   \begin{table}
\begin{center}
\begin{threeparttable}
\caption[]{A summary of the color measurements of (297274) 1996 SK 
on May 22 and 23}\label{tab3}
 \begin{tabular}{cccccc}
  \hline\noalign{\smallskip}
$UT_V$ &V	&B-V &V-R &V-I &$Airmass^*$\\
  \hline\noalign{\smallskip}
May 22& & & & & \\
13:30:36& $16.259\pm0.006$ & $0.840\pm0.012$ & $0.520\pm0.007$ & $0.769\pm0.008$ & 1.553\\
13:31:58& $16.270\pm0.006$ & $0.835\pm0.012$ & $0.511\pm0.007$ & $0.769\pm0.008$ & 1.546\\
14:20:46& $16.390\pm0.007$ & $0.652\pm0.013$ & $0.459\pm0.009$ & $0.847\pm0.009$ & 1.373\\
14:22:10& $16.386\pm0.006$ & $0.644\pm0.012$ & $0.475\pm0.008$ & $0.859\pm0.008$ & 1.370\\
15:07:03& $16.063\pm0.005$ & $0.748\pm0.010$ & $0.472\pm0.007$ & $0.827\pm0.007$ & 1.297\\
15:08:26& $16.065\pm0.005$ & $0.741\pm0.010$ & $0.478\pm0.007$ & $0.832\pm0.007$ & 1.296\\
  \hline\noalign{\smallskip}
May 23& & & & & \\
17:34:31 & $16.495\pm0.007$ & $0.906\pm0.017$ & $0.457\pm0.009$ & $0.707\pm0.012$ & 1.523\\
  \hline\noalign{\smallskip}
Mean & $16.275\pm0.016$ & $0.767\pm0.033$ & $0.482\pm0.021$ & $0.801\pm0.023$ & \\
  \hline\noalign{\smallskip}
\end{tabular}
%% place \tablecomments and \tablerefs below \end{center| and \end{center}:
%% you may leave the table-width parameter to editors or set to your actual size
\begin{tablenotes}
\item *: The airmass is displayed for the time of V-band observed because the BVRI observations in each color measurement were obtained in sequential order in a short time interval within 11 min.
   \end{tablenotes}
   \end{threeparttable}
   \end{center}
\end{table}

\begin{figure}
   \centering
   \subfigure{\label{fig2a}
   \includegraphics[width=0.6\columnwidth]{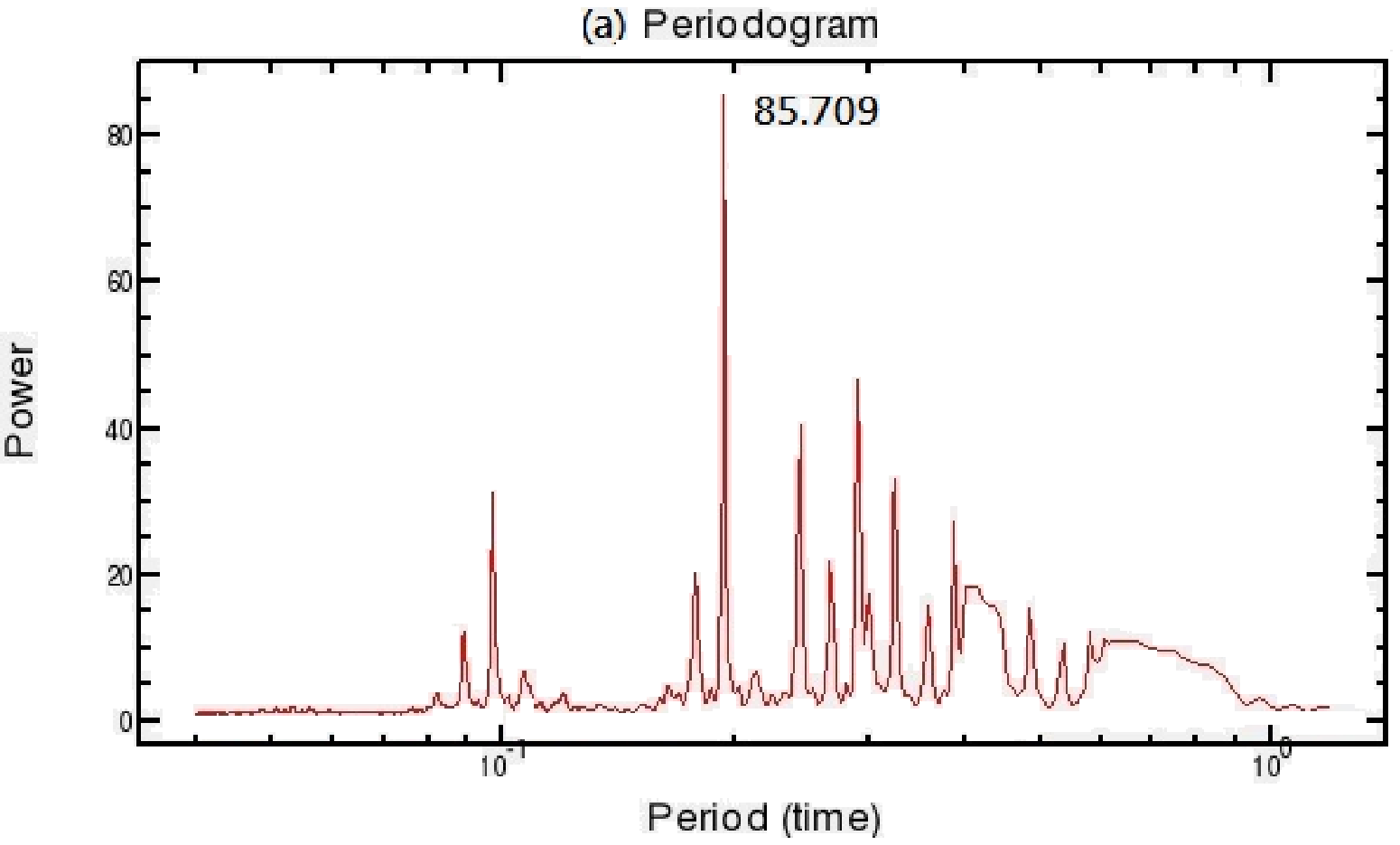}
   }
    \subfigure{\label{fig2b}
   \includegraphics[width=0.8\columnwidth]{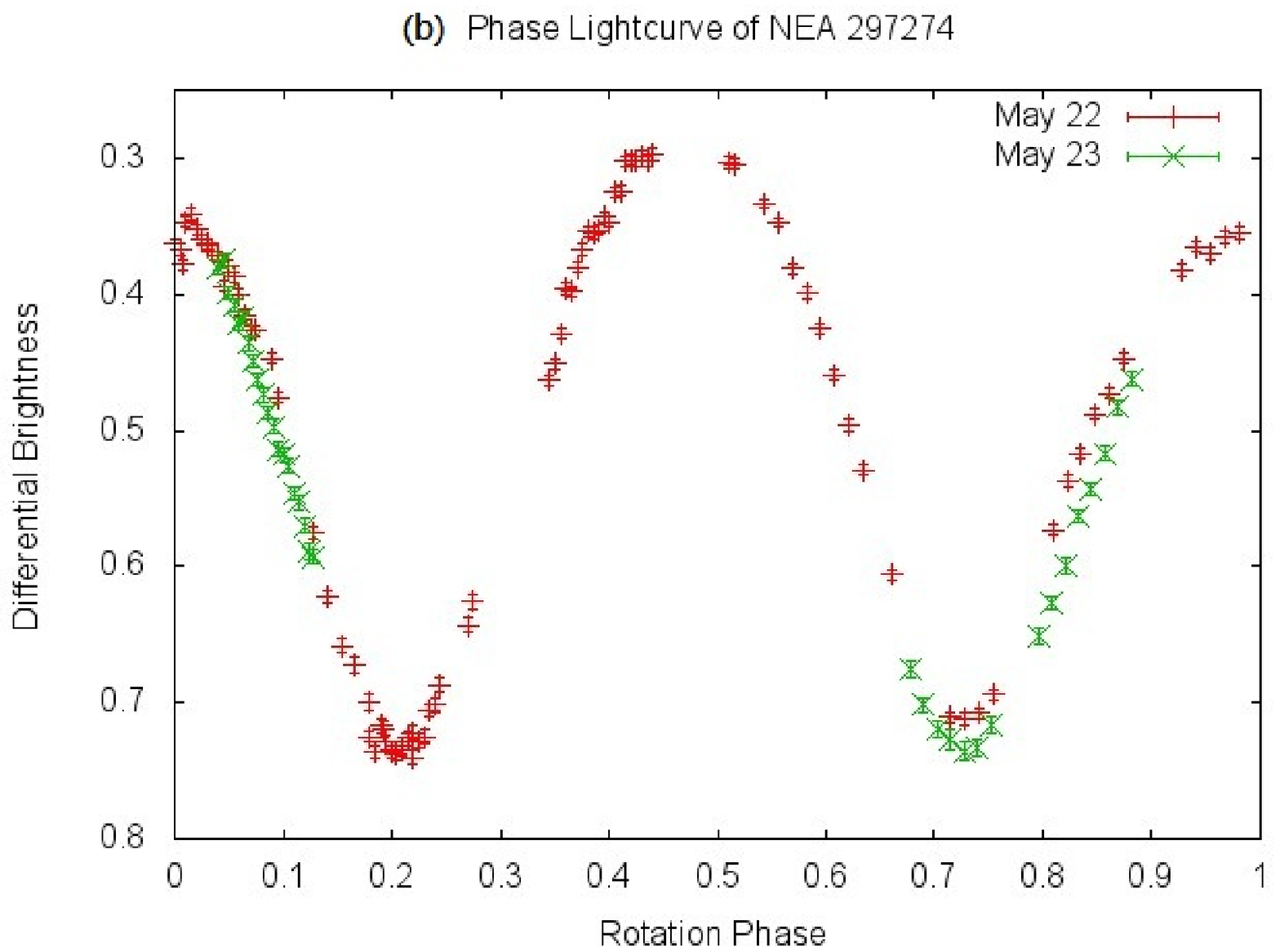}
   }
   
   \caption{The periodogram (a) and phase lightcurve (b) of (297274) 1996 SK consisting of both the data on May 22 and 23. The differential brightness is normalized to be consistent for two days.}
   \label{Fig2}
   \end{figure}

Using the Plavchan algorithm (Plavchan et al.~\cite{Pla08}) to compute periodogram, the spin period of (297274) 1996 SK was found to be $4.656\pm0.122$ hours. The uncertainty in the frequency was estimated based on the method of Horne et al.~(\cite{Hor86}). The periodogram and the folded lightcurve from the R-band measurements along with the rotation phase are shown in Figure~\ref{Fig2}. The lightcurve shows that (297274) 1996 SK has a rather smooth configuration. For an ellipsoidal shape of the asteroid, the peak-to-peak variation ($\Delta m$) of the lightcurve can be used to calculate the ratio of the long axis to short axis ($a/b$) according to the formula $\Delta m = 2.5log(a/b)$. From the lightcurve of the (297274) 1996 SK, the $\Delta m$ was 0.44. It means that the $a/b$ is about 1.50. However, since the above a/b value is obtained by assuming that the asteroid was observed at an aspect angle (i.e., the angle between the line of sight and spin axis) of $90^\circ$, the actual axial ratio ($a/b$) may be more than that. The asteroid’s diameter ($D$) can be calculated by using the formula (Yoshida et al.~\cite{Yos04}), $logD=3.130-0.5logA-0.2H$, where $H$ is the absolute magnitude and $A$ is the surface albedo (Yoshida et al., 2004). Assuming $A = 0.2$ (corresponding to the mean albedo of S-type asteroids) and $H = 16.866$ mag for (297274) 1996 SK, its diameter is 1.28 km. The long axis and the short axis can be computed to be 1.57 km and 1.05 km, respectively.

\begin{figure}
   \centering
   \includegraphics[width=0.9\columnwidth]{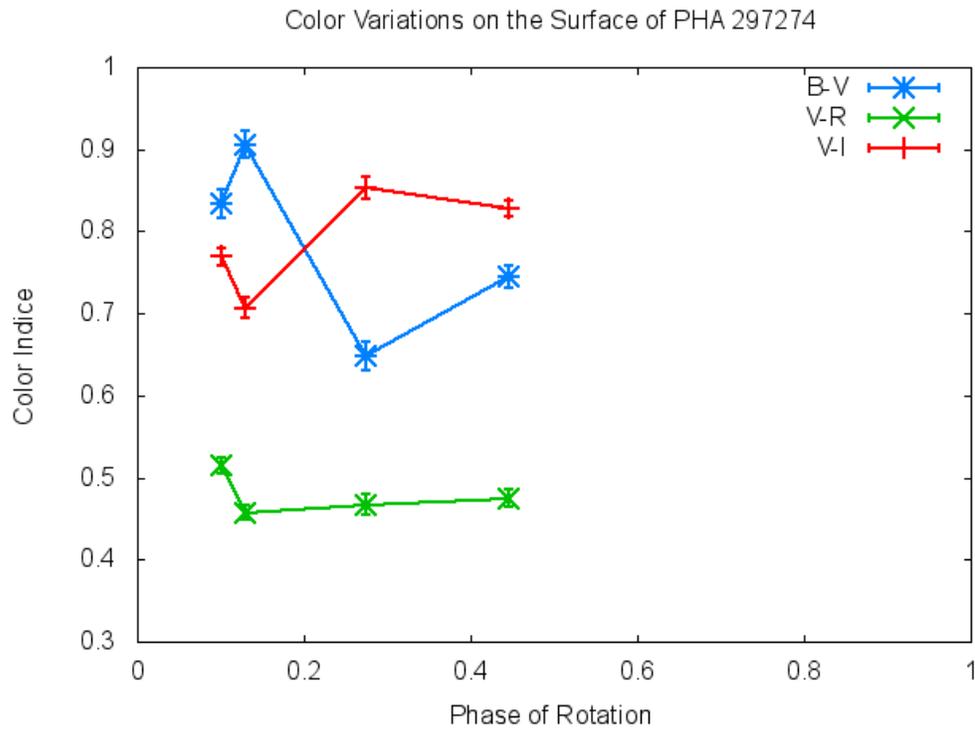}
   \caption{Surface color variations of PHA (297274) 1996 SK over the phase of rotation.}
   \label{Fig3}
   \end{figure}

Table~\ref{tab3} summarizes the color measurement results obtained on March 22 and 23. The multi-wavelength observations of (297274) 1996 SK at several different times allow us to estimate the color indices and to examine the possible changes of its surface color during the rotation. Figure~\ref{Fig3} displays the color variations at four phases of rotation observed on the two days. It reveals that both B-V and V-I colors vary significantly, while V-R change is comparatively small. The maximum changes between the phase 0 to 0.5 for B-V, V-R and V-I are 0.258, 0.058 and 0.146, respectively. Such a large range of the color variation indicates the possible presence of surface heterogeneity on (297274) 1996 SK. 

\begin{figure}[H]
   \centering
   \includegraphics[width=0.9\columnwidth]{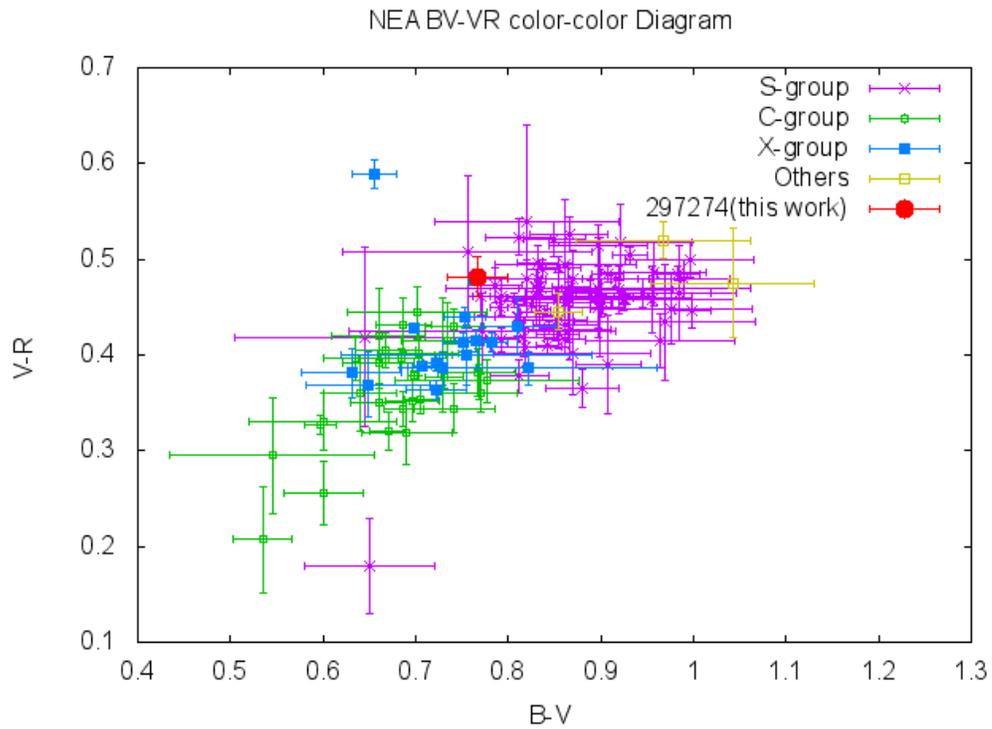}
   \caption{The color-color diagram of NEAs with known color indices in different taxonomic types [$\alpha$] and (297274) 1996 SK observed from this work.($\alpha$: Betzler et al., 2010; Carbognani, 2008; Dandy et al., 2003; Hapke, 2000; Hergenrother et al., 2009; Hicks et al., 2011, 2012, 2013; Jewitt et al., 2006, 2013; Karashevich et al., 2012; Pieters et al., 2000; Ye, 2011.)}
   \label{Fig4}
   \end{figure}

The brightness magnitudes of the B, V and I bands follow the general trend of the R-band lightcurve. The average values of $B-V = 0.767\pm 0.016$, $V-R = 0.482\pm 0.021$, and $V-I = 0.801\pm 0.025$ of (297274) 1996 SK can be compared with the known colors from different taxonomies of NEAs determined by previous observations archived in the "Data Base of Physical and Dynamical Properties of NEAs" from European Asteroid Research Node. These results are plotted in Figure~\ref{Fig4}, which shows the B-V and V-R terms generally divided into S-group (S, Q, R-types et al.) , X-group (X, E-types et al.) and C-group (C, F, B-types et al.) of NEAs. It indicates that the surface color of (297274) 1996 SK locates on the boundary between S-group and X-group asteroids.

   \begin{figure}
   \centering
   \includegraphics[width=0.9\columnwidth]{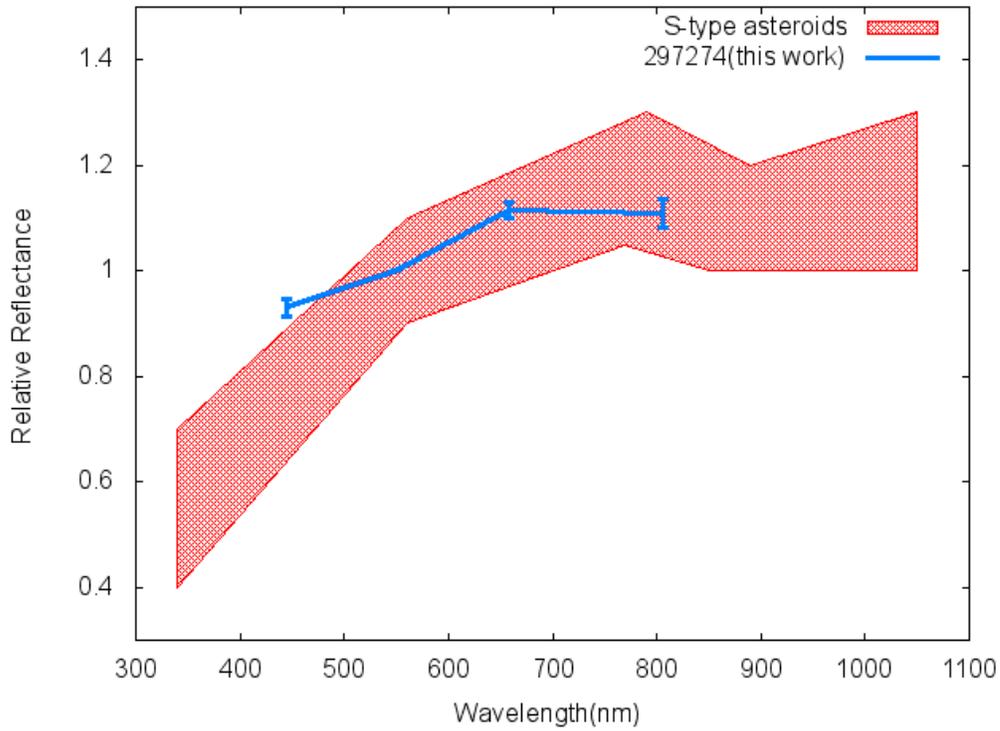}
   \caption{Relative reflectance spectrum of (297472) 1996 SK from our data (thick line) in comparison with those integrated spectra of S-type asteroids from the archived data of “Small Bodies Node”. The shaded area indicates the range of the S-type spectra.}
   \label{Fig5}
   \end{figure}
   
   \begin{figure}[H]
   \centering
   \includegraphics[width=0.9\columnwidth]{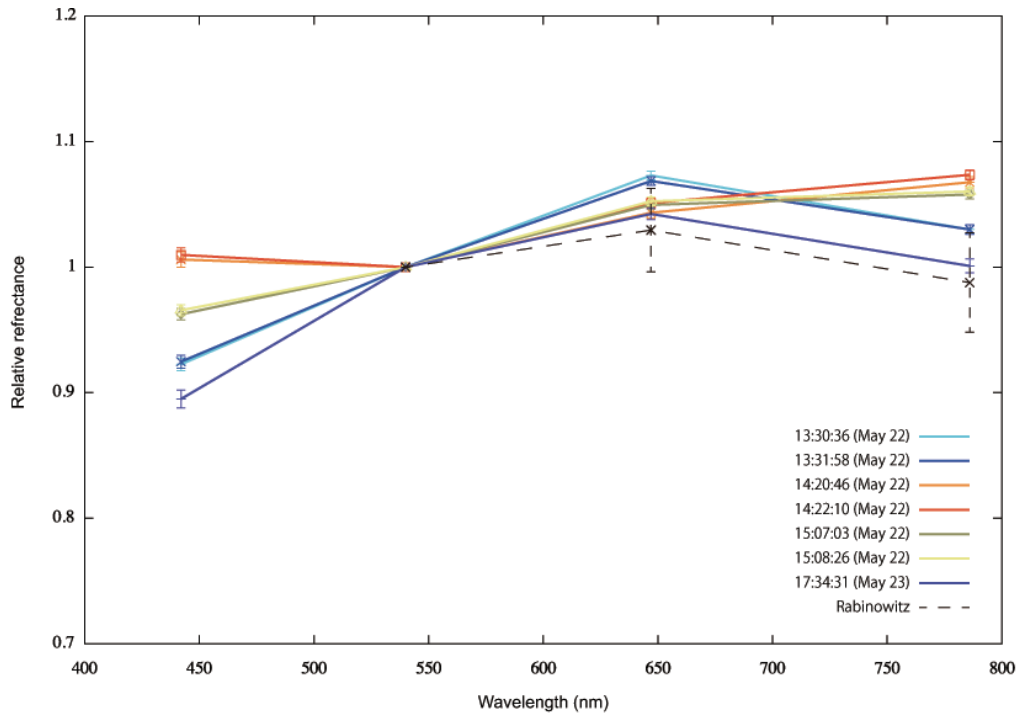}
   \caption{A comparison of the relative reflectance spectra of (297274) 1996 SK taken at different times of March 22 and 23 with that reported by Rabinowitz (1998).}
   \label{Fig6}
   \end{figure}

Figure~\ref{Fig5} illustrates the average relative reflectance spectrum of (297274) 1996 SK obtained by subtracting the solar colors $B-V = 0.665$, $V-R = 0.367$, and $V-I = 0.705$ (Howell,~\cite{How95}) from its colors. It falls into the spectral region of the S-type asteroids, so (297274) 1996 SK should be classified as a member of S-type objects. It is interesting to note that Rabinowitz (\cite{Rab98}) reported color measurements of (297274) 1996 SK in October, 1996, with $V-R = 0.430\pm 0.070$ and $V-I = 0.678\pm 0.0587$, respectively. These values are close to the corresponding results obtained on May 23 (see Table~\ref{tab3} and Figure~\ref{Fig6}) similar to shape of Q-type asteroids while still having significant differences from those taken at other times. The possible implication will be discussed later.

\section{Summary and Discussion}
\label{sect:summary&discussion}

Our observations of PHA (297274) 1996 SK at opposition in May, 2012, lead to the following conclusion:

\begin{enumerate}
\item The rotation period of this asteroid is found to be $4.656\pm 0.122$ hours, i.e., well below the spin cutoff of 2.2 hours. 
\item The amplitude of lightcurve variability is $\Delta m = 0.44$ indicating an elongated shape with the ratio of the long axis to the short axis $(a/b) = 1.50$ but possibly underestimated. 
\item The average color indices of $B-V = 0.767\pm 0.033$, $V-R = 0.482\pm 0.021$, $V-I = 0.801\pm 0.025$ and the corresponding surface reflectance means that (297274) 1996 SK belongs to the S-type taxonomic class. With the surface albedo assumed to be 0.2, which is typical value of S-type asteroids, and $H_v = 16.866$, the projected long and short axes are 1.57 km and 1.05 km, respectively.
\item Over the rotation range of 133 degrees, (297274) 1996 SK displays significant color changes which might imply the existence of large change in mineralogical and/or compositional variation on its surface.
\end{enumerate}

The detection of large color change is an important result of this work because it could mean that (297274) 1996 SK might contain various properties of its surface spectra. Because there is no information on the relation of the color measurements to the rotational phase in the work of Rabinowitz (\cite{Rab98}), it is difficult to analyze the cause of the color differences between our present results and his work. One thing is nearly certain; that is they could not be caused by short-term space weathering effect since the associated time scale is at least on the order of a million years (Vernazza et al.~\cite{Ver09}). From this point of view, the existence of inhomogeneous surface composition or differential space weathering effect would be the most viable explanation. The first scenario would mean that (297274) 1996 SK might contain the interface material of some differentiated region of its parent body at impact disruption. The second scenario has been discussed by Yoshida et al (\cite{Yos04}) in the case of the color variation of (832) Karin - see also Sasaki, T. et al. (\cite{Sas04, Sas06a}), Sasaki, S. et al. (\cite{Sas06b}), and Ito and Yoshida (\cite{Ito07}). This could have come about by micrometeoroid impact process on young and older surface areas (Clark et al.~\cite{Cla02}; Sasaki et al.~\cite{Sas04}). Figure 6 shows the relative reflectance spectra observed in different time varying from S-type to Q-type. It might be also related to the second scenario that the asteroid has two parts of weathered and un-weathered surfaces. It was a possibility that Rabinoiwtz (\cite{Rab98}) had measured the colors of the vicinity of the phase which we observed on May 23. Both possibilities mean that (297274) 1996 SK should not be covered by a homogeneous regolith layer of small particles. Could this surface cleansing be achieved by tidal breakup in previous close encounters with Earth and other terrestrial planets as proposed by Nesvorny et al (\cite{Nes10})? These are issues we plan to investigate in the future work. 

\begin{acknowledgements}
This work was partially supported by the Project 019/2010/A2 of Macau Science and Technology Development Fund: MSAR No. 0166 and Taiwan Ministry of Education under the Aim for Top University Program NCU. The Lulin Observatory is operated by Institute of Astronomy, National Central University, Taiwan, under a grant of NSC 96-2752-M-008-011-PAE.
\end{acknowledgements}

\label{lastpage}

\end{document}